# Zero-field spin-orbit-torque switching driven by magnetic spin Hall effect

Po-Hung Lin[1], Po-Wei Lee[1], Yu-Hsuan Lin[2], Bo-Yuan Yang[1], Vinod Kumar[1], Hsiu-Hau Lin[2]*, Chih-Huang Lai[1]*

[1]Department of Materials Science and Engineering, National Tsing Hua University, Hsinchu, 300, Taiwan

[2]Department of Physics, National Tsing Hua University, Hsinchu, 300, Taiwan

**Abstract**

Spin Hall effect plays an essential role in generating spin current from the injected charge current, following the Dyakonov–Perel rule that the directions of charge flow, spin flow and spin polarization are mutually perpendicular to each other. Recently, its generalization from an antiferromagnet, so-called magnetic spin Hall effect, has been studied and verified by measuring anomalous spin accumulations. Here, we investigate the magnetic spin Hall effect in bilayer materials made of a heavy metal and an antiferromagnet. The spin current generated by the magnetic spin Hall effect accomplishes spin-orbit-torque switching for ferromagnetic magnetization and exchange bias concurrently without any external magnetic field. The switching mechanism crucially relies on the non-collinear spin texture in the antiferromagnet, capable of generating symmetry-breaking components in the spin-current tensor so that the external magnetic field is no longer necessary. The zero-field concurrent switching of magnetization and exchange bias is a significant technological breakthrough. Furthermore, our findings pave the way to explore the magnetic spin Hall effects in various spin textures through spin-orbit-torque switching.

*Correspondence and requests for materials should be addressed to

H.-H. Lin email: hsiuhau.lin@gmail.com

C.-H. Lai email: chlai@mx.nthu.edu.tw

**Introduction.** Spin Hall effect (SHE), converting charge current into spin current, plays an essential role in modern spintronics[1,2]. Due to spin-orbit interactions, the unpolarized charge current is scattered into perpendicular direction so that the charge flow, spin flow and spin polarization are mutually perpendicular. Because detecting spin current directly is challenging, it is often measured through spin accumulations on the sample edges[3] or the spin-orbit torque (SOT) switching of neighboring magnetization[4–9].

The SOT switching is a promising technique to manipulate the magnetization direction of magnetic tunnel junctions with high switching speed and less damage to tunneling barrel between the top and bottom ferromagnetic layers[10]. However, there is a serious setback for the SOT switching driven by spin Hall effect. In usual experimental setup, the device is made of multilayers, containing heavy metal (HM) and ferromagnet (FM). When a charge current pulse passes through the device, a spin current in the perpendicular direction is generated by the HM layer and flows into the ferromagnetic layer. Consider the case where the magnetization of the ferromagnetic layer has strong perpendicular magnetic anisotropy (PMA). During the current pulse, the generated SOT pushes the perpendicular magnetic moments into the in-plane configurations.[8,11] Without symmetry-breaking interactions, these in-plane magnetic moments relax to the perpendicular axis randomly after the current pulse. In consequence, the SOT drives the ferromagnetic layer into a demagnetization state or multi-domain state.

To facilitate a definite SOT switching of magnetization, a longitudinal magnetic field (along the current direction) is needed to break the symmetry. During the current pulse, the push-down magnetic moments are not strictly in-plane anymore. The longitudinal field breaks the symmetry so that the magnetic moments are slightly tilted, making the SOT switching definite. The necessity of an external field renders the SOT switching unpractical for realistic applications. To remove the required longitudinal field, there have been several proposals to achieve field-free SOT switching for magnetization, including wedge structure[12-14], in-plane exchange bias[15-17], multiple spin-current sources[18,19] and so on. These attempts try to replace the longitudinal field by some other symmetry-breaking mechanisms so that the field-free SOT switching can be achieved.

But, one can also tackle the challenge from a more fundamental perspective. Note that, while spin current is often described as spins with definite polarization flows in the specific direction, it is actually a rank-two tensor with multiple components.

This tensor property of spin current has been demonstrated by detecting anomalous spin accumulations in recent experiments under the name of magnetic spin Hall effect (mSHE)[20]. This breakthrough is inspiring and provides a strong hint that the zero-field SOT switching may be achievable by properly designed mSHE.

Following the inspiring thread, we fabricate the bilayer structure composed of HM and antiferromagnet (AFM) as the spin-current source. When the charge current injected into the Pt/FeMn bilayer, the spin current, following the usual Dyakonov–Perel rule, is first generated in the Pt layer. The spin current then flows into FeMn layer and gets scattered by the Neel order with non-collinear spin texture. It is rather remarkable that the resultant spin current (with additional non-vanishing components) can achieve zero-field SOT switching. We will explain this surprising outcome in the following paragraphs.

**Concurrent zero-field SOT switching.** We first demonstrate the experimental findings for the zero-field SOT switching of magnetization and exchange bias simultaneously. The structure of the sample used here is substrate/Ti (3)/Cu (6)/[Co (0.3)/Ni (0.6)]$_2$/Co (0.3)/Ni (0.3) /FeMn (8)/Pt (5)/Ti (2), where the layer thickness is denoted by the number in the parentheses (unit = nm). The substrate is thermally oxidized silicon. All films are deposited by dc magnetron sputtering. The bottom and top Ti layers are adhesion and capping layers, respectively, and the Cu layer is needed to maintain PMA in the ferromagnetic Co/Ni layers. The spin current is generated in the Pt layer first and then get modified by the FeMn antiferromagnet.

The film is made into single wire device of $10\,\mu m \times 10\,\mu m$ size by photolithography and ion-beam etching. The electrode is done by lift-off process. As shown in Figure 1, the SOT switching curve is taken by applying current pulses of $10\,\mu s$ width, and the magnetization is measured by focus magneto-optical Kerr effect (FMOKE). Without any external magnetic field, as shown in Figure 1(c), the ferromagnetic Co/Ni layers can be switched when the current pulse exceeding the threshold. After the SOT switching, the hysteresis-loop measurement (Figure 1(d)(e)) manifests the reversal of exchange bias.

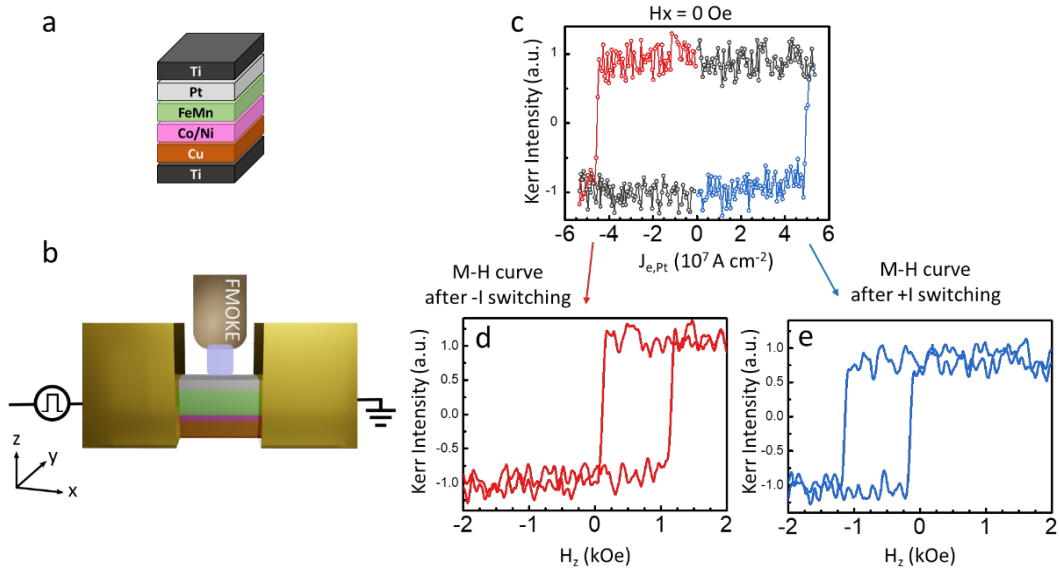

**Figure 1.** Field-free SOT switching and the concurrent reversal of exchange bias. (a) The film structure. (b) The measurement setup. (c) The field-free SOT switching curve. (d)(e) The M-H curve measured after each SOT switching.

To avoid confusions in later discussions, it is necessary to define the polarity of the SOT switching. When -M (initial) can be flipped to +M (final) by positive current density, the polarity of the SOT switching is referred as "positive". Meanwhile, positive polarity also indicates that +M (initial) can be flipped to -M (final) by large enough negative current density. It is straightforward to define the negative polarity in the similar fashion. Following the above definitions, it shall be clear that the SOT switching in Figure 1 carries positive polarity.

The concurrent zero-field SOT switching is unique in several ways. The first surprise is of course the zero-field switching. In conventional SOT switching, the longitudinal magnetic field (along the charge current direction) breaks the $Z_2$ symmetry in the perpendicular axis and sets the polarity of the SOT switching. When the external field is reversed, the polarity turns opposite as well. In the absence of the external field, the $Z_2$ symmetry is retained and the conventional mechanism for SOT switching fails. In the FM/AFM/HM trilayer structure, the external field is not needed to achieve the SOT switching, hinting that there exists symmetry-breaking mechanism escaping our reasoning at this point.

The second surprise is the concurrent reversal of exchange bias. It has been recently shown that the exchange bias can be manipulated by spin-orbit torque in HM/FM/AFM trilayer structure[21–23]. Here the sample structure is revered yet the concurrent exchange-bias switching is observed. Our findings here serve as optimistic evidence that the exchange bias can be manipulated by spin-orbit torque in wide variety of magnetic multilayers.

The last but not the least surprise is the long penetration length of the spin current. It seems that the spin current, generated by the Pt layer, travels through the 8nm-thick FeMn antiferromagnetic layer and remains strong enough to flip the ferromagnet underneath. The zero-field SOT switching is also observed in samples where the thickness of the AFM layer is 10 nm (not shown here). It has been reported in recent works that the spin current can persist through a relatively thick AFM or ferrimagnetic layer[24,25]. The mechanisms include spin-wave propagation in AFM insulator[26] or the compensated spin precession due to opposite spin arrangements in the Neel order[27]. Our finding agrees with previous reports in the literature but the underlying mechanism, elaborated in later paragraphs, may not be exactly the same. Note that FeMn is a metallic antiferromagnet and the persistence of spin current over such a long distance is truly surprising.

**Requisite of the AFM/HM bilayer.** As will become clear later, the observed zero-field SOT switching is achieved by the exquisite combination of the AFM/HM bilayer: the HM layer provides the major spin current (and thus the spin-orbit torque) and the AFM layer scatters the injected spin current and generates the symmetry-breaking torque. We will first demonstrate the necessity to have both layers, followed by the detail theoretical analysis showing the resultant spin current through the bilayer.

To establish the requisite of the AFM/HM bilayer, we design two samples without Pt and FeMn layers respectively. The structure of the sample without Pt layer is substrate/Ti 3/Cu 6/[Co/Ni]$_3$ 2.4/ FeMn 8/Ti 2 (unit = nm), referred as null-Pt sample. As shown in Figure 2, no trace of SOT switching is found in the null-Pt sample, suggesting the Pt layer is the dominant source of the spin current. This is consistent with previous studies, showing the spin Hall effect in FeMn is negligible.[28]

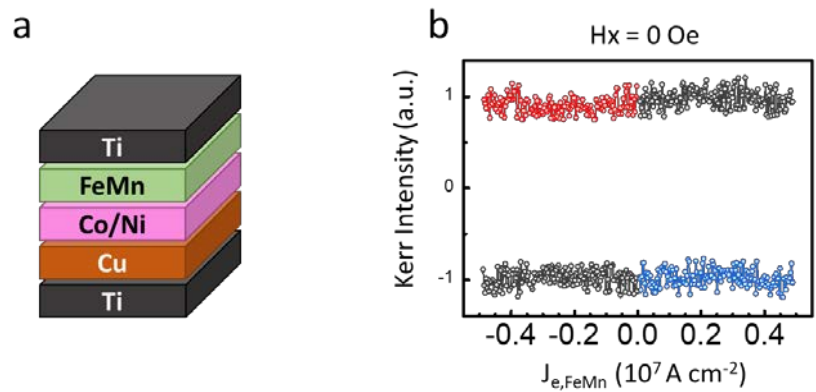

**Figure 2.** The switching curve for null-Pt sample at Hx = 0 Oe.

The sample structure without FeMn layer is substrate/Ti 3/Cu 6/[Co/Ni]$_3$ 2.4/ Pt 5/Ti 2 (unit = nm), referred as null-FeMn sample. Without external magnetic field, as shown in Figure 3(a), the SOT created by the current pulse drives the ferromagnet into demagnetization state. This transition is anticipated because, without symmetry-breaking interactions, the magnetic moments relax back into the perpendicular axis with random directions, leading to the demagnetization state (or multi-domain state).

The SOT switching is recovered when the external field in present. As shown in Figure 3(b)(c), the polarities of the SOT switching in the presence of Hx = +100 and -100 Oe are opposite, well explained by conventional SOT switching mechanism.

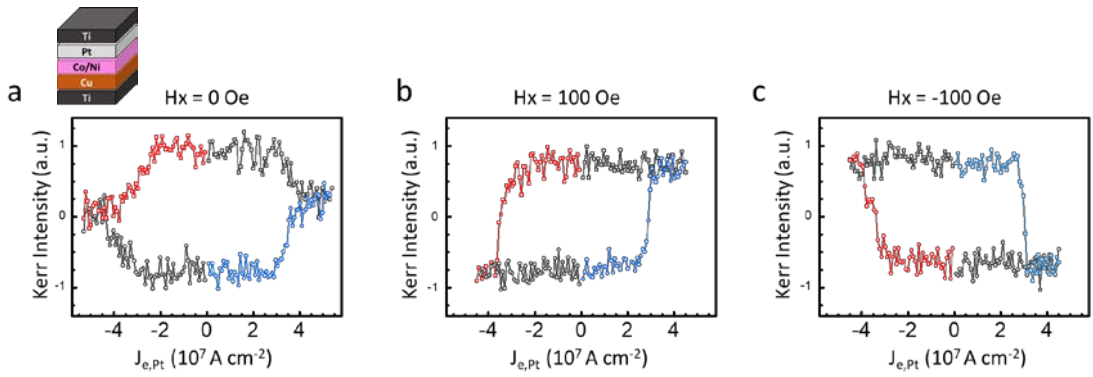

**Figure 3.** The SOT switching curve for null-FeMn sample. (a) Hx = 0 Oe. (b) Hx = 100 Oe (c) Hx = -100 Oe. The pulse width is 10μs.

From the null-Pt and null-FeMn samples, we show that the AFM/HM bilayer plays an essential role for achieving the concurrent field-free SOT switching. It is tempting to suggest that the spin current is generated by the Pt layer and received modification through the FeMn layer by the Neel order with non-collinear spin texture. But, one needs to explain how the symmetry-breaking torque is generated and also why the spin current can survive while passing through the thick FeMn layer. Surprisingly, all these puzzles can be answered by carefully analyzing the interaction between the spin current (generated by the Pt layer) and the Neel order with non-collinear spin texture (in the FeMn layer).

**AFM scattered spin current.** To study the interaction between the spin current and the Neel order, it is important to recall that the spin current is a rank-two tensor $J_i^\alpha$, where the subindex $i$ (Roman) represents the spatial direction and the superindex $\alpha$ (Greek) represents the spin orientation. When the charge current passes through the Pt layer, a spin current $J_i^\alpha = \frac{\hbar}{2e} J_c \theta_{\text{SH}} \times \delta_{iz}\delta_{\alpha y}$ is generated[20], where the spin polarization is along $y$ axis and the spin current flows along $z$ direction into the FeMn layer.

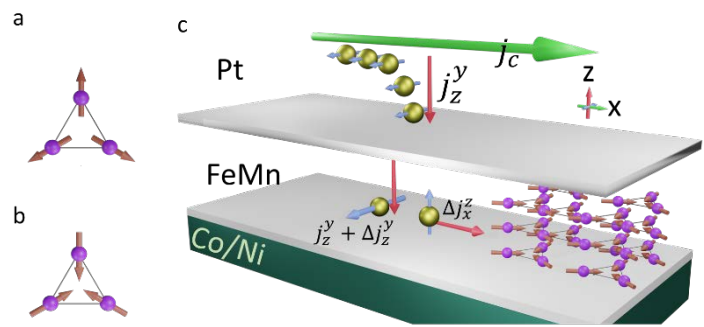

**Figure 4.** When the charge current flows through the Pt/FeMn bilayer, non-trivial components of the spin-current tensor is generated due to magnetic SHE. (a) and (b)

show the spin textures of FeMn with opposite Neel order.

The spin current is then scattered by the Neel order with non-collinear spin texture. The detail calculations can be found in Supplementary Information and the resultant spin current is $J'^{\alpha}_i = J^{\alpha}_i + \Delta J^{\alpha}_i$, where the correction to the original spin current is characterized by a rank-two tensor $\Delta J^{\alpha}_i$. Because AFM is polycrystalline, the interaction between the spin current $J^{\alpha}_i$ and the Neel order cannot be computed by the usual first-principles calculations. Instead, one needs to seek for scattering matrix caused by the microscopic spin texture. Here we assume the interactions between the itinerant spins (spin current) and the localized spins (Neel order) take the form of Heisenberg exchange coupling. After computing the $T$-matrix from solving the Lippmann-Schwinger equation, the scattered spin-current tensor $\Delta J^{\alpha}_i$ is presented in the matrix form below:

$$\Delta J^{\alpha}_i = \begin{pmatrix} \Delta J^x_x & \Delta J^y_x & \Delta J^z_x \\ \Delta J^x_y & \Delta J^y_y & \Delta J^z_y \\ \Delta J^x_z & \Delta J^y_z & \Delta J^z_z \end{pmatrix} = \begin{pmatrix} a_\parallel \cos\varphi_N & a_\parallel \sin\varphi_N & A_\parallel \cos\varphi_N \\ -a_\parallel \sin\varphi_N & a_\parallel \cos\varphi_N & -A_\parallel \sin\varphi_N \\ 0 & A_z & 0 \end{pmatrix}$$

where $\varphi_N$ denotes the orientation of the non-collinear spin pattern within the unit cell (Figure S1 in SI). The coefficient functions can be classified into two groups by their strengths: the capitalized $A_z$ and $A_\parallel$ represent large components in the spin-current tensor while the lower-case $a_\parallel$ represents small components. In addition to the original component $J^y_z$ in the injected spin current, other components arise due to scattering with the non-collinear spin texture. In general, the scattered spin current $\Delta J_i^\alpha$ would depend on the crystal structure and the spin arrangement. But, after averaging over the scattered angles, it can be shown that the scattered spin current $\Delta J_i^\alpha$ only depends on the angle $\varphi_N$ of the Neel order.

It is important to classify the symmetries of these coefficient functions. It turns out that both major coefficients $A_z$ and $A_\parallel$ reverse signs (odd symmetry) when the current pulse is reversed, while the minor coefficient $a_\parallel$ remains the same (even symmetry). Because the minor coefficient is relatively small and carries even symmetry (insensitive to current reversal), we would neglect its contribution in the following discussions.

Let us explain why the spin current flowing in the $z$ direction does not decay significantly in AFM. Note that $\Delta J_z^x = 0 = \Delta J_z^z$ so that the only non-zero component flowing along the z axis is $\Delta J_z^y = A_z$. That is to say, the scattered spin current flows into the ferromagnetic layer carries the same spin orientation as the original one generated by the Pt layer. According to the $T$-matrix calculations, the major coefficient $A_z$ is positive in the weak scattering regime. In consequence, after scattered by the non-collinear spin texture, the spin-current component $J'^y_z = J^y_z + \Delta J^y_z$ is enhanced. Of course, there are other scattering mechanisms (like non-magnetic defects) leading to diffusive dynamics. Thus, the calculations presented here do not mean that the spin current will grow exponentially. Instead, it provides a reasonable explanation why the spin current will not attenuate significantly when penetrating through thick (8-10 nm) antiferromagnet FeMn.

The zero-field switching can be explained by the non-vanishing components in the scattered spin current. To achieve definite SOT switching, the resultant torque must satisfy the right symmetries. Take the SOT switching with positive polarity as an example. When the initial magnetization is set in +M direction, only the negative current pulse will cause the SOT switching while the positive current pulse keeps the magnetization intact. Therefore, the switching torque should carry odd symmetry when the current pulse is reversed. On the other hand, keeping the current pulse positive, only the −M configuration can be switched while the +M configuration remains intact. Thus, the switching torque should carry odd symmetry when the magnetization is reversed.

The in-plane components of the spin current $A_\parallel$ obey the right symmetry to achieve SOT switching. Due to magnetic interactions, the non-collinear spin arrangement is slightly canted[29]. So, when the magnetization of the FM layer is flipped, the Neel order near the AFM/FM interface is also flipped due to strong interfacial exchange coupling between FM and AFM. The flipped Neel order corresponds to the change of $\varphi_N \to \varphi_N + \pi$. Because the sinusoidal functions change signs with $\pi$ shift, the $A_\parallel$ components carry odd symmetry when the FM magnetization is reversed. In addition, as evident in Equation (S23) in Supplementary Information, the $A_\parallel$ components also changes signs when the current direction is reversed. The in-plane spin current $\Delta J_i^z = A_\parallel (\cos \varphi_N, -\sin \varphi_N, 0)$ thus carries the odd symmetries upon current and magnetization reversals. The resultant torque $\tau^z = -\nabla \cdot \Delta J_i^z = -\partial_x \Delta J_x^z - \partial_y \Delta J_y^z$ thus has the right symmetries required by the SOT switching.

Neither the sign of the perpendicular $\Delta J_z^y$ nor the symmetries of the in-plane $\Delta J_i^z$ components of the spin current is fine-tuned in our model. Once the interactions between the injected spin current and the non-collinear spin texture are properly included, the zero-field SOT switching through a thick antiferromagnet can be explained naturally. Summarizing our theoretical analyses above, the composite material of the HM/AFM bilayer can be viewed as a unique source of spin current with non-trivial mSHE, capable of driving the zero-field SOT switching of magnetization and exchange bias simultaneously.

**Polarity change of the SOT switching.** The polarity of the zero-field SOT switching can be changed in various ways. The easier approach is to replace the top Pt layer with Ta. As shown in Figure 5, the key layer structure consists of Cu 6/[Co/Ni]$_3$ 2.4/ FeMn 8/Ta 2 (unit = nm), showing zero-field SOT switching with negative polarity. Due to the opposite sign of the spin-Hall angle in the Ta layer, as oppose to the original Pt layer, the coefficient function $A_\parallel$ in the AFM layer also changes signs so that the polarity alters from positive to negative.

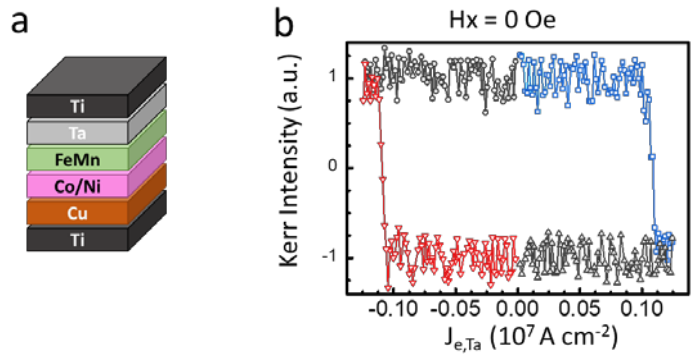

**Figure 5.** The field-free switching for top-Ta sample. (a) The film structure (b) The field-free SOT switching with negative polarity. Because the spin-Hall angle $\theta_{SH}$ is negative in the Ta layer, the polarity of the zero-field SOT switching changes to negative, as oppose to positive for the top-Pt sample.

**Discussions.** It may be tempting to explain the zero-field SOT switching by an emergent effective field generated by the Neel order during the current pulse. Just as the Neel order scatters the spin current and gives rise to non-trivial corrections, the spin current can change the spin configurations in the Neel order as well. However, the symmetry criteria seem to rule out the possibility. Assume that the effective longitudinal field is generated by the spin current flowing through the Neel order. When the current is reversed, the effective field should also change to the opposite direction. This odd symmetry upon current reversal ruins the definite polarity of the SOT switching and is in direct conflict with our experimental findings.

Previously, we demonstrate how the polarity of the zero-field SOT switching can be changed by replacing the HM layer. At this point, it is not easy to determine the polarity because the simple model contains parameters with unknown signs. But, if one can modify the distribution of the Neel-order domains in AFM, it is expected that the polarity change can occur. Thus, we applied the in-situ magnetic field (created by a permanent magnet on the sample holder) during deposition for the multilayer structure. Note that the deposition chamber has a +Hz stray field without putting the permanent magnet due to the configuration of the magnetron sputtering system. To change distribution of the Neel-order domains in AFM, we deposited the films with in-situ Hz = -800 Oe, referred to the -Hz sample. The zero-field SOT switching of the –Hz sample is shown in Figure 6. It is clear that the polarity of the -Hz sample changes to negative. By simply reversing the in-situ magnetic field during deposition, the polarity of the SOT switching can be reversed. It seems that the polarity is closely related to the distribution of the Neel-order domains in the FeMn layer but further investigations are needed to manifest the connection between the polarity and the spin texture in the AFM layer.

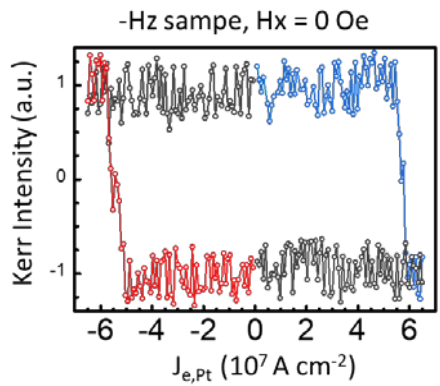

**Figure 6.** The zero-field SOT switching for the -Hz sample.

In conclusions, we investigate the magnetic spin Hall effect in bilayer materials made of a heavy metal and a non-collinear antiferromagnet. The symmetry-breaking torque is automatically generated by the in-plane components of the scattered spin current. It is astounding that the zero-field SOT switching of magnetization and exchange bias can be achieved by the spin current arisen from mSHE of the HM/AFM bilayer. Our findings bridge the intimate connection between mSHE and the zero-field SOT switching, opening up fresh approaches to manipulate magnetization and exchange bias via spin-current source with sophisticated spin textures.

# Supplementary Information

## S1. Scattered spin current in noncollinear AFM

Previous theories have dealt with how charge current scatters with impurities through spin-orbit coupling and given rise to spin current, commonly known as the spin Hall effect. In our experiment, a theory of how spin current scatters with spin texture is needed to explain zero-field switching in AFM. We consider electrons scatter with an equilateral triangular impurity consisting of three spins located at the vertices of the triangle with zero net magnetization. We assume that the electrons scatter with the three vertices within a triangle coherently, and the scattering events between different unit cells are incoherent. The above process gives rise to various components of scattered spin currents (Eq. (S15)) and provides critical symmetry-breaking spin current for zero-field switching in AFM.

**Interaction between itinerant spin and spin texture.** We apply pseudopotential approximation and model the potential in each equilateral triangular unit in AFM with Heisenberg interaction:

$$V(\vec{r}) = J \sum_{i=1}^{3} (\vec{\sigma} \cdot \vec{s}_i) \, \delta(\vec{r} - \vec{R}_i) \quad (S1)$$

In the above equation, $J$ is the coupling constant between the itinerant spin ($\vec{\sigma}$) and the localized spins ($\vec{s}_i$) at each vertex of the triangular unit; $\vec{R}_i$ is the position of the three vertices where the origin is set at the center of the triangle. $\vec{s}_i$ and $\vec{R}_i$ are described by two azimuthal angles $\varphi_S$ and $\varphi_R$:

$$\begin{cases} \vec{s}_1 = (\cos\varphi_s, \sin\varphi_s, 0) \\ \vec{s}_2 = (\cos(\varphi_s + 2\pi/3), \sin(\varphi_s + 2\pi/3), 0) \quad (S2) \\ \vec{s}_3 = (\cos(\varphi_s + 4\pi/3), \sin(\varphi_s + 4\pi/3), 0) \end{cases}$$

$$\begin{cases} \vec{R}_1 = \dfrac{l}{\sqrt{3}}(\cos\varphi_R, \sin\varphi_R, 0) \\ \vec{R}_2 = \dfrac{l}{\sqrt{3}}(\cos(\varphi_R + 2\pi/3), \sin(\varphi_R + 2\pi/3), 0) \quad (S3) \\ \vec{R}_3 = \dfrac{l}{\sqrt{3}}(\cos(\varphi_R + 4\pi/3), \sin(\varphi_R + 4\pi/3), 0) \end{cases}$$

Here, $l$ is the length of the equilateral triangular unit, and we only consider localized spins ($\vec{s}_i$) in xy-plane. As shown in Figure S1, the orientations of the three spins are labeled by $\varphi_s$, $\varphi_s + 2\pi/3$, $\varphi_s + 4\pi/3$, while their positions at the vertices of the crystal triangle are labeled by $\varphi_R$, $\varphi_R + 2\pi/3$, $\varphi_R + 4\pi/3$. Because the injected spin current from the HM layer is polarized in y-direction, so the z-component of $\vec{s}_i$ leads to higher order correction of scattered spin currents, ignored in the calculations here. In the following discussion, the localized spins are quasi-static and are treated classically, while we treat the itinerant spin with quantum dynamics.

**Lippmann-Schwinger equation formalism.** To understand the scattered spin currents in AFM, one needs to solve the scattered wave after colliding with an impurity according to the Lippmann-Schwinger equation:

$$\langle \vec{r}|\Psi_{\vec{p}}\rangle = \langle \vec{r}|\phi_{\vec{k}\sigma}\rangle + \int d^3\vec{r}'\, g_R(\vec{r},\vec{r}')\langle \vec{r}'|V|\Psi_{\vec{p}}\rangle \quad (S4)$$

$$= \langle \vec{r}|\phi_{\vec{k}\sigma}\rangle + J\sum_{i=1}^{3} g_R(|\vec{r} - \vec{R}_i|)(\vec{\sigma}\cdot\vec{s}_i)\langle \vec{R}_i|\Psi_{\vec{p}}\rangle \quad (S5)$$

In the above equation, $\langle \vec{r}|\phi_{\vec{k}\sigma}\rangle$ is the free propagating wave:

$$\langle \vec{r}|\phi_{\vec{k}\sigma}\rangle = e^{i\vec{k}\cdot\vec{r}}\chi_\sigma \quad (S6)$$

where $\chi_\sigma$ is the spinor state. $\mathcal{G}_R(\vec{r},\vec{r}')$ is defined as the unperturbed retarded Green's function:

$$\mathcal{G}_R(\vec{r},\vec{r}') = \left\langle \vec{r} \left| \frac{1}{E_k - \hat{H}_0 + i0^+} \right| \vec{r}' \right\rangle = -\frac{m}{2\pi\hbar^2} \frac{e^{ik|\vec{r}-\vec{r}'|}}{|\vec{r}-\vec{r}'|} = \mathcal{G}_R(|\vec{r}-\vec{r}'|) \quad (S7)$$

**Calculation of T-matrix.** The scattering amplitude from $|\phi_{\vec{k}\sigma}\rangle$ to $|\phi_{\vec{p}\sigma'}\rangle$ is characterized by T-matrix:

$$T_{\vec{p}\vec{k}}(\sigma',\sigma) \equiv \langle \phi_{\vec{p}\sigma'} | V | \Psi_{\vec{p}} \rangle \quad (S8)$$

$$= \sum_{i=1}^{3} \langle \phi_{\vec{p}\sigma'} | \vec{R}_i \rangle J(\vec{\sigma} \cdot \vec{s}_i) \langle \vec{R}_i | \Psi_{\vec{p}} \rangle \quad (S9)$$

In the above equation, $\langle \vec{R}_i | \Psi_{\vec{p}} \rangle$ can be solved by inserting $\vec{r} = \vec{R}_i$ in the Lippmann-Schwinger equation:

$$\begin{cases} \langle \vec{R}_1 | \Psi_{\vec{p}} \rangle = \langle \vec{R}_1 | \phi_{\vec{k}\sigma} \rangle + J\mathcal{G}_R(0)(\vec{\sigma} \cdot \vec{s}_1)\langle \vec{R}_1 | \Psi_{\vec{p}} \rangle + J\mathcal{G}_R(0)[(\vec{\sigma} \cdot \vec{s}_2)\langle \vec{R}_2 | \Psi_{\vec{p}} \rangle + (\vec{\sigma} \cdot \vec{s}_3)\langle \vec{R}_3 | \Psi_{\vec{p}} \rangle] \\ \langle \vec{R}_2 | \Psi_{\vec{p}} \rangle = \langle \vec{R}_2 | \phi_{\vec{k}\sigma} \rangle + J\mathcal{G}_R(0)(\vec{\sigma} \cdot \vec{s}_2)\langle \vec{R}_2 | \Psi_{\vec{p}} \rangle + J\mathcal{G}_R(0)[(\vec{\sigma} \cdot \vec{s}_3)\langle \vec{R}_3 | \Psi_{\vec{p}} \rangle + (\vec{\sigma} \cdot \vec{s}_1)\langle \vec{R}_1 | \Psi_{\vec{p}} \rangle] \\ \langle \vec{R}_3 | \Psi_{\vec{p}} \rangle = \langle \vec{R}_3 | \phi_{\vec{k}\sigma} \rangle + J\mathcal{G}_R(0)(\vec{\sigma} \cdot \vec{s}_3)\langle \vec{R}_3 | \Psi_{\vec{p}} \rangle + J\mathcal{G}_R(0)[(\vec{\sigma} \cdot \vec{s}_1)\langle \vec{R}_1 | \Psi_{\vec{p}} \rangle + (\vec{\sigma} \cdot \vec{s}_2)\langle \vec{R}_2 | \Psi_{\vec{p}} \rangle] \end{cases} \quad (S10)$$

Here, the Green's function at the origin ($\mathcal{G}_R(0)$) requires an introduction of cut-off momentum $k_c \sim a_0^{-1} \gg k$ in order to resolve the divergence behavior, where $a_0$ is the characteristic radius of the orbital:

$$\mathcal{G}_R(0) = \frac{2m}{\hbar^2} \frac{1}{(2\pi)^3} \int d^3\vec{k}' \frac{1}{k^2 - k'^2 + i0^+} = -\frac{m}{\pi^2\hbar^2}\left(a_0^{-1} + \frac{\pi i k}{2}\right) \quad (S11)$$

Next, we arrive at the T-matrix by inserting the solutions of $\langle \vec{R}_i | \Psi_{\vec{p}} \rangle$ into Eq. (S9):

$$T_{\vec{p}\vec{k}}(\sigma',\sigma)|_{\vec{k}=k\hat{z}}$$

$$= \chi_{\sigma'}^{\dagger} \frac{\begin{pmatrix} J^2(\mathcal{G}_R(0) - \mathcal{G}_R(l))\sum_{i=1}^{3} e^{-i\Delta_j} & Je^{-i\varphi_s}\left(e^{-i\Delta_1} + e^{-i\frac{2\pi}{3}}e^{-i\Delta_2} + e^{i\frac{2\pi}{3}}e^{-i\Delta_3}\right) \\ Je^{-i\varphi_s}\left(e^{-i\Delta_1} + e^{i\frac{2\pi}{3}}e^{-i\Delta_2} + e^{-i\frac{2\pi}{3}}e^{-i\Delta_3}\right) & J^2(\mathcal{G}_R(0) - \mathcal{G}_R(l))\sum_{i=1}^{3} e^{-i\Delta_j} \end{pmatrix}}{1 - J^2(\mathcal{G}_R(0)^2 + \mathcal{G}_R(l)\mathcal{G}_R(0) - 2\mathcal{G}_R(l)^2)} \chi_\sigma \quad (S12)$$

In the above equation, $\Delta_j \equiv \vec{p} \cdot \vec{R}_j$. Note that the T-matrix presented above is for incident wave travelling in $+z$ direction ($\vec{k} = k\hat{z}$).

**Scattered spin currents.** The scatter spin currents ($\Delta J_i^\alpha$) in AFM can be modeled by the T-matrix given by Eq.(S12). An incident spin current $J_z^y$ injected into AFM can be viewed as electrons with spin polarized in $+y$ direction ($-y$ direction) travelling in $+z$ direction into AFM if $J_z^y > 0$ ($J_z^y < 0$). As a result, the scattered spin currents are given by the following:

$$\Delta \vec{J}^\alpha(J_z^y > 0) = C \int_{\text{forward}} d\Omega_p \, k(\sin\theta_p \cos\varphi_p, \sin\theta_p \sin\varphi_p, \cos\theta_p)$$
$$\left[\left|T_{\vec{p}\vec{k}}(+\alpha, +y)\right|^2 - \left|T_{\vec{p}\vec{k}}(-\alpha, +y)\right|^2\right] \quad (S13)$$

$$\Delta \vec{J}^\alpha(J_z^y < 0) = C \int_{\text{forward}} d\Omega_p \, k(\sin\theta_p \cos\varphi_p, \sin\theta_p \sin\varphi_p, \cos\theta_p)$$
$$\left[\left|T_{\vec{p}\vec{k}}(+\alpha, -y)\right|^2 - \left|T_{\vec{p}\vec{k}}(-\alpha, -y)\right|^2\right] \quad (S14)$$

In the above equations, $\Delta \vec{J}^\alpha \equiv (\Delta J_x^\alpha, \Delta J_y^\alpha, \Delta J_z^\alpha)$, and we consider elastic scattering ($|\vec{k}| = |\vec{p}| = k$); $C$ is a positive constant that relates differential cross section to $\left|T_{\vec{p}\vec{k}}(\sigma', \sigma)\right|^2$ and is independent of the scattering angle and incident momentum $k$. Next, we arrive at the scattered spin currents by inserting Eq. (S12) into Eqs. (S13) and (S14):

$$\Delta J_i^\alpha = \begin{pmatrix} \Delta J_x^x & \Delta J_x^y & \Delta J_x^z \\ \Delta J_y^x & \Delta J_y^y & \Delta J_y^z \\ \Delta J_z^x & \Delta J_z^y & \Delta J_z^z \end{pmatrix} = \begin{pmatrix} a_\parallel \cos\varphi_N & a_\parallel \sin\varphi_N & A_\parallel \cos\varphi_N \\ -a_\parallel \sin\varphi_N & a_\parallel \cos\varphi_N & -A_\parallel \sin\varphi_N \\ 0 & A_z & 0 \end{pmatrix} \quad (S15)$$

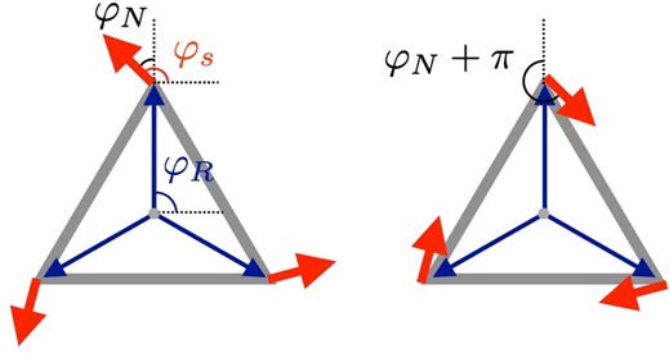

Figure S1. Non-collinear spin texture for the Neel order $\varphi_N$ and $\varphi_N + \pi$.

Here, the Neel order can be characterized by $\varphi_N \equiv \varphi_s - \varphi_R$ (see Figure S1), and the constants $a_\parallel$, $A_\parallel$, and $A_z$ are given by the following:

$$a_\parallel(+y) = \frac{6\sqrt{3}Ck\pi J^3(\sin kl - kl\cos kl)}{|1 - J^2(\mathscr{g}_R(0)^2 + \mathscr{g}_R(l)\mathscr{g}_R(0) - 2\mathscr{g}_R(l)^2)|^2 k^2 l^2}\text{Im}[\mathscr{g}_R(l) - \mathscr{g}_R(0)] \quad (S16)$$

$$a_\parallel(-y) = a_\parallel(+y) \quad (S17)$$

$$A_\parallel(+y) = \frac{6\sqrt{3}Ck\pi J^3(\sin kl - kl\cos kl)}{|1 - J^2(\mathscr{g}_R(0)^2 + \mathscr{g}_R(l)\mathscr{g}_R(0) - 2\mathscr{g}_R(l)^2)|^2 k^2 l^2}\text{Re}[\mathscr{g}_R(0) - \mathscr{g}_R(l)] \quad (S18)$$

$$A_\parallel(-y) = -A_\parallel(+y) \quad (S19)$$

$$A_z(+y) = \frac{3Ck\pi J^4|\mathscr{g}_R(0) - \mathscr{g}_R(l)|^2\left(1 + \frac{4J_1(kl)}{kl}\right)}{|1 - J^2(\mathscr{g}_R(0)^2 + \mathscr{g}_R(l)\mathscr{g}_R(0) - 2\mathscr{g}_R(l)^2)|^2} \quad (S20)$$

$$A_z(-y) = -A_z(+y) \quad (S21)$$

Here, $J_1(x)$ is the Bessel function of the first kind (not to be confused with the exchange coupling $J$, or the spin-current tensor $J_i^\alpha$), and $+y$ ($-y$) corresponds to the polarization of the incident spin current into AFM. In the weak scattering limit $ka_0 \ll kl \ll 1$ and $|J\mathscr{g}_R(0)| \ll 1$, the above coefficients are simplified as follow

$$a_{\parallel}(+y) = a_{\parallel}(-y) \approx \frac{\sqrt{3}CmJ^3k^5l^3}{6\hbar^2} \propto J^3k^5l^3 \quad (S22)$$

$$A_{\parallel}(+y) = -A_{\parallel}(-y) \approx -\frac{2\sqrt{3}CmJ^3k^2la_0^{-1}}{\pi\hbar^2} \propto J^3k^2la_0^{-1} \quad (S23)$$

$$A_z(+y) = -A_z(-y) \approx \frac{9Cm^2J^4ka_0^{-2}}{\pi^3\hbar^4} \propto J^4ka_0^{-2} \quad (S24)$$

According to the above equations, the relative magnitude of the constants $a_{\parallel}$, $A_{\parallel}$, and $A_z$ are given by the following:

$$\frac{a_{\parallel}}{A_{\parallel}} \sim \mathscr{O}\left(k^3l^3\frac{a_0}{l}\right) \ll 1 \quad (S25)$$

$$\frac{A_z}{A_{\parallel}} \sim \mathscr{O}\left(\frac{J g_R(0)}{kl}\right) \quad (S26)$$

From Eqs. (S25) and (S26), the dominant scattered spin currents for spin current $J_z^y$ injected into AFM are $\Delta J_x^z$, $\Delta J_y^z$, and $\Delta J_z^y$.

**Limitation.** The above calculation is applicable for polycrystalline AFM samples. In polycrystalline AFM samples, the scattering of itinerant electrons with different unit cells in AFM are incoherent. Therefore, we are allowed to treat each unit cell in AFM as an impurity. On the other hand, for single-crystalline AFM, one has to consider itinerant electrons scattering with the whole sample coherently and the first-principles calculations can deliver more accurate descriptions. Another limitation is that the above calculation is carried out within a simple parabolic band. Inclusion of realistic band structures can improve the results but the qualitative trends shall remain robust.